\documentclass[12pt]{article}
\usepackage{amsmath,amssymb,amsfonts,amsthm}
\usepackage{graphicx}
\usepackage{epsfig,multicol}

\newcommand{\be}[0]{\begin{equation}}
\newcommand{\ee}[0]{\end{equation}}
\newcommand{\ba}[0]{\begin{eqnarray}}
\newcommand{\ea}[0]{\end{eqnarray}}

\title{Hard-gluon Evolution of Nucleon Generalized Parton Distributions in Soft-Wall AdS/QCD Model
}

 \author{ Majid Dehghani
\\
\footnotesize{Particle Physics Group, Faculty  of Physics,
Yazd University, Yazd, Iran}
\\ \footnotesize{e-mail:m.dehghani56@stu.yazd.ac.ir  } }
\begin{document}

 \date{\today}
\maketitle

{\bf Keywords:\footnotesize  Holographic soft-wall model;Generalized Parton Distributions;DGLAP equations;AdS/QCD}

{\bf PACS:} 13.40.Gp, 14.20.Dh, 13.60.Fz


 \newcommand{\norm}[1]{\left\Vert#1\right\Vert}
 \newcommand{\abs}[1]{\left\vert#1\right\vert}
 \newcommand{\set}[1]{\left\{#1\right\}}
 \newcommand{\R}{\mathbb R}
 \newcommand{\I}{\mathbb{I}}
 \newcommand{\C}{\mathbb C}
 \newcommand{\eps}{\varepsilon}
 \newcommand{\To}{\longrightarrow}
 \newcommand{\BX}{\mathbf{B}(X)}
 \newcommand{\HH}{\mathfrak{H}}
 \newcommand{\D}{\mathcal{D}}
 \newcommand{\N}{\mathcal{N}}
 \newcommand{\la}{\lambda}
 \newcommand{\af}{a^{ }_F}
 \newcommand{\afd}{a^\dag_F}
 \newcommand{\afy}{a^{ }_{F^{-1}}}
 \newcommand{\afdy}{a^\dag_{F^{-1}}}
 \newcommand{\fn}{\phi^{ }_n}
 \newcommand{\HD}{\hat{\mathcal{H}}}
 \newcommand{\HDD}{\mathcal{H}}

\abstract{Holographic soft-wall model is successful in the phenomenology of hadrons. Here with the use of generalized parton distributions (GPDs) obtained from AdS/QCD, perturbative effects are entered into the formalism. Perturbations are incorporated in the formalism through the evolution of GPDs according to the DGLAP like equations. Evolved proton GPDs are compared with a phenomenological model to show that we can get good improvements of the holographic model. It seems that combining the holographic soft-wall model with perturbative effects to some extent, gives the correct physics of GPDs. }

 \section{Introduction}\label{sec-intro}
 Quantum chromodynamics (QCD) is the accepted fundamental theory of strong interactions. Perturbative QCD is
successful in the phenomenology of hadron physics at high energies. But it is not yet possible to describe
hadrons in terms of its quark and gluon degrees of freedom.
Perturbative calculations in quantum field theory (QFT) is based on the mathematical formulation of Feynman diagrams, and many observables of theory are in agreement with these calculations.
Continuation of perturbative calculations to low energies (large distances) is not possible due to the increase in
 coupling constant and breakdown of perturbation theory. There are also other indications that we have to deal with nonperturbative phenomena in QCD. Several methods to handle non-perturbative quantum field theories have been invented and this active field has witnessed an enormous progress.  But there are still challenging open problems such as confinement, and determination of each parton species contribution to hadron characteristics such as mass and spin.

For many years nonperturbative methods in QFT were at the front line to solve QCD dilemmas.
In recent years another idea from string theory, the AdS/CFT correspondence \cite{maldacena1999large}, is used as a differnt tool in describing strongly coupled field theories.
The AdS/CFT formalism in its original form can't be applied to real QCD. As will be explained in next section, it is the modified form that can be applied to QCD, the so called AdS/QCD models. Two main divisions of AdS/QCD models are: top-down approaches in which D-brane constructions are used, and bottom-up approaches where builds the model in a 5-dimensional space that contains gravity.

The importance of bottom-up approaches among the other AdS/QCD models lies in their phenomenological success. Bottom-up approaches are based on the existence of a local effective action on AdS background that reproduces desired properties of QCD in the 4D boundary theory. The holographic soft-wall model is in this category with some good physical and mathematical features, that will be concerned here. Holographic soft-wall model is a semiclassical first approximation to QCD \cite{de2012hadronic}. In the boundary gauge theory, perturbations are important and causes many observable effects. Some examples are running of coupling constant, running of effective mass and DGLAP equations in deep inelastic scattering (DIS) formalism. Although of a different nature, running of coupling constant is automatically incorporated in holographic soft-wall model, as will be explained in Sec.\ref{gpdevo}.
Here the implications of another perturbative effect, the DGLAP equations, will be considered.  As will be explained in Sec.\ref{gpdevo}, DGLAP equations result from the fact that we need the PDFs to be independent of scale parameter. Scale parameter separates perturbative and nonperturbative scales.  The same role as PDFs in DIS formalism, is played by generalized parton distribution functions (GPDs) in exclusive processes such as deeply virtual Compton scattering (DVCS). In an exclusive process also, there is a scale parameter that separates perturbative and nonperturbative regimes. scale independence of nonperturbative part, also results in DGLAP like equations in exclusive events.

From the phenomenological point of view also, there are different observables to probe the nucleon content. The most famous observables
are structure functions and form factors that unravel the momentum distribution and charge distribution of hadrons, respectively.
Both approaches are complementary, but they are not appropriate ones for a nonperturbative treatment of hadron structure.
Generalized parton distributions (GPDs) are unique tools to describe hadrons in terms of its fundamental degrees of freedom, i.e. quarks and gluons\cite{belitsky2005unraveling,diehl2003generalized}. GPDs relate different classes of physical observables and collect them in a single framework. The above mentioned ideas and their combination, have been successful in describing some aspects of strongly interacting particles. We will elaborate on how perturbative corrections enter into the formalism of soft-wall model.

The idea of this work is taken from DIS formalism, where perturbative application of QCD, results in the evolution of  parton distribution functions (PDFs) according to DGLAP equations.
Here the DGLAP like equations are applied to GPDs obtained in holographic models. GPDs in holographic soft-wall and hard-wall models are obtained for the first time in Refs. \cite{vega2011generalized} and \cite{vega2012generalized}, respectively.
Here the GPDs of holographic soft-wall model \cite{vega2011generalized}  are  evolved according to DGLAP like equations. Evolved GPDs incorporate the perturbative effects and can be used for a direct comparison with experimental data or relate them to hadronic observables through sum rules. We use a simpler method, that is to compare the results with the GPDs of a phenomenological model. Because phenomenological models are arranged to fit experimental data, this is a simple and effective method to see how much we can get closer to data by evolved quantities.
To do perturbative calculations, we need the gauge theory side coupling constant, that is the dual of the gravity side. Light-front holography is used to obtain the physical coupling constant of the gauge theory side in soft-wall model. Then the GPDs are evolved according to DGLAP like equations for valence quarks and discussing the implications. It will be shown that many features of proton GPDs are contained in the combination of holographic soft-wall model with perturbative effects.

The organization of the paper is as follows: section \textbf{\ref{swmodel}} contains a brief survey of holographic soft-wall model and the form factors of nucleons which are the tools to obtain GPDs. GPDs have a central role in our perturbative calculations, so they will be introduced in section \textbf{\ref{gpdsw}} along with a phenomenological model. Section \textbf{\ref{gpdevo}} is the main part of the report, first light-front holography is used to obtain the coupling constant of boundary gauge theory and then GPDs are evolved according to DGLAP equations. Finally, the resulting changes of proton GPDs are compared with a phenomenological model.


\section{Holographic Soft-Wall Model and Nucleon Form Factor }\label{swmodel}

  Beginning of AdS/CFT conjecture was a correspondence  between type IIB string theory and $N=4$ supersymmetric Yang-Mills(SYM) theory in the large $N_c$ limit \cite{gubser1998gauge,maldacena1999large}.
$N=4$ SYM theory is a conformal theory and the AdS metric on gravity side is also conformal. Real QCD in not a conformal theory so the original AdS metric is modified in AdS/QCD models  to take into account the nonconformality.
AdS/QCD models connect theories living in higher dimensional AdS space to 4-dimensional QCD-like theories. Top-down AdS/QCD models  \cite{sakai2005low,klebanov2000supergravity,babington2004chiral} are based on finding D-brane configurations with features similar to QCD, where bottom-up models\cite{erlich2005qcd,da2005chiral,hirn2005interpolating},  rely on the existence of a local effective action on AdS background that reproduces the desired properties of QCD in the 4-dimensional boundary theory.

Holographic soft-wall model is an important member of bottom-up approaches and is the base of this work.
As explained above, we need to break conformal invariance in order to incorporate confinement in the formalism, and there are different ways to do it. In holographic soft-wall model, breaking of conformal invariance is based on the introduction of additional warp factor or, equivalently a dilaton background, on the 5D gravity side.

The metric of soft-wall model restricts the interactions to a limited region of AdS space, so it is a built-in confined theory.
In this model a dilaton  profile of the form $e^{\pm\kappa^2 z^2} $ results in a monotonic potential.
In the soft-wall model, for the boson and fermion states of 4D gauge theory one introduces explicit bosonic and fermionic fields in the 5D lagrangian, respectively.
The terms in the 5D Lagrangian are chosen based on simplicity, symmetries and its relevance to the problem\cite{abidin2009nucleon}.
Then AdS/QCD duality relates normalizable solutions of 5D gravity side to the states of the 4D boundary gauge theory.

Nucleons are particles of spin-1/2 and in the present approach an explicit fermion field in five dimensional AdS space needs to be introduced.
Fermions in a soft-wall model for nucleon form factors, was first introduced by Abidin and Carlson\cite{abidin2009nucleon}. We continue with a brief introduction to the formalism of soft-wall model.
The five dimensional space of the gravity side is an anti-de Sitter space with the metric in Poincare (conformal) coordinate without additional warp factor of the form:
  \begin{equation}\label{ads metr}
     ds^2=g_{MN} dx^M dx^N=\frac{1}{z^2}\eta_{\mu\nu} dx^\mu dx^\nu - dz^2
       \end{equation}
where $\eta_{\mu\nu}=diag(1,-1,-1,-1)$, $\mu\nu=0,1,2,3$ and $z$ the holographic radial coordinate extended from zero to $\infty$ with the boundary at $z=0$.
The soft-wall model infrared cutoff results from the introduction of an exponential dilaton field.
Here notice that, as discussed in \cite{de2012hadronic}, the essential physics in holographic soft-wall model does't depend on the dilaton profile sign and that they can be related to each other by a field redefinition. Results of the positive and negative profile sign differ from each other by a surface term.
But as will be discussed in Sec.\ref{gpdevo}, to have a physical coupling constant for the gauge theory side, a positive sign dilaton profile is preferred. So we conveniently use either of two signs, except in the discussions of coupling constant.
Inserting an overall exponential dilaton field in the formalism leads to an infrared cutoff for bosons, but it doesn't work for fermions.
 This was resolved by adding a dilaton interaction to the mass term \cite{abidin2009nucleon}.

The matching procedure used in \cite{vega2011generalized} to obtain the GPDs of holographic soft-wall model, for the first time. The same procedure also applied in the context of holographic hard-wall model \cite{vega2012generalized}.
Here we only outline the method used in \cite{vega2011generalized}. Here we write the nucleon form factors in the soft-wall model obtained in \cite{abidin2009nucleon}:
 \begin{equation}\label{f1f2}
\begin{aligned}
   F_1^p(Q^2)&=C_1(Q^2)+\eta_p C_2(Q^2)   \\
   F_2^p(Q^2)&=\eta_p C_3(Q^2) \\
   \\
    F_1^n(Q^2)&=\eta_n C_2(Q^2) \\
    F_2^n(Q^2)&=\eta_n C_3(Q^2) \\
    \end{aligned}
 \end{equation}
where $Q^2=-t$ and $C_i(Q^2)$ are defined in App.(\ref{appena}):
expressing these functions in an integral form is the tool in establishing the connection with GPDs.
This was a  brief introduction to holographic soft-wall model. In our approach the perturbative effects are entered into the formalism through the generalised parton distributions (GPDs). So in the next section we introduce the GPDs and outline the way they are obtained in soft-wall formalism \cite{vega2011generalized}.

\section {Generalized Parton Distributions}\label{gpdsw}

  From perturbative QCD formalism we know that parton distribution functions (PDFs) are appropriate tools in formulation of deep inelastic scattering (DIS) observables, such as structure functions. Structure functions in deep inelastic scattering (DIS) processes, are the most important observables in probing the nucleon content \cite{belitsky2005unraveling}. Deep inelastic scattering is an inclusive reaction, so limited information is contained in it.
 In exclusive processes kinematical parameters of all initial and final particles are detected and so contain much more information. GPDs are entered into the description of exclusive deep inelastic scatterings, such as deeply virtual Compton scattering (DVCS) \cite{diehl2003generalized}.

For exclusive processes, similar to the inclusive DIS case, factorization to long and short distances is valid. Factorization scheme will be explained in next section, here we only note that in the factorized form, the long distance nonperturbative parts are GPDs.
Mathematically GPDs are defined as matrix elements of the same operator as PDFs, except between states with different momenta\cite{burkardt2003impact}.
For twist-2 operators there exist two kinds of helicity independent GPDs. The GPDs for valence quarks in the nucleon are $H^q(x,\xi,t)$ and $E^q(x,\xi,t)$, where $t=q^2=-Q^2$ is momentum transfer squared, $\xi$ is skewness and $x$ the light cone momentum fraction. As a consequence of Lorentz invariance, the $\xi$ dependence of GPDs drops out in the sum rule for the form factors of quark vector and axial vector currents  \cite{diehl2005generalized} , so our discussion will be restricted to the case: $\xi=0$ and changing the notation to the form: $H^q(x,\xi,t)\rightarrow H^q(x,t) $.
 Now quoting some useful limits of GPDs that we need for the description of hadronic properties later.
 In the forward limit GPDs reduce to PDFs \cite{diehl2003generalized}:
    \begin{align}\label{pdf}
  H^q(x,0,0)&=q(x)  \;\;\;\;\;\ for\;\ x>0 \nonumber \\
  H^q(x,0,0)&=-\bar{q}(x) \;\;\ for\;\ x<0
   \end{align}
where $q(x)$ and $\bar{q}(x)$ are distribution functions for quarks and and anti-quarks, respectively.
with the above mentioned properties, for the valence quark we have \cite{guidal2005nucleon}:
\begin{equation}\label{valen}
 H_v^q(x,t)=H^q(x,t)+H^q(-x,t)
 \end{equation}
 This is the combination entering the proton and neutron dirac form factors. In analogy to above equation the valence GPDs $E_v^q(x,t)$, are introduced as:
 \begin{equation}\label{valenE}
 E_v^q(x,t)=E^q(x,t)+E^q(-x,t)
 \end{equation}

  Note that the forward limit($t=0$) of the GPDs $E_v^u(x,t)$, that gives rise to $F_2$ structure function of nucleons cannot be expressed in terms of any known parton densities\cite{guidal2005nucleon}.
  There are sum rules that relate the integral of GPDs to the form factors \cite{guidal2005nucleon}.
  In  \cite{vega2011generalized} using the integral representation of form factors and sum rules that relate them to GPDs the following expressions for distribution functions in the soft-wall formalism is obtained:

 \begin{eqnarray}\label{Hqsw}
 H_v^q(x,Q^2)=q(x)x^a  \nonumber \\
 E_v^q(x,Q^2)=\varepsilon^q(x)x^a
 \end{eqnarray}
where $a=\frac{Q^2}{4\kappa^2}$; and see App.(\ref{appenb}) for functional forms of $q(x)$ and $\varepsilon^q(x)$.

The numerical value of $\kappa$ depends on the concerned  physical observable. In \cite{abidin2009nucleon} the $\kappa=350MeV$ value is obtained from the requirement that it gives the nucleon masses. But for other observables  such as hadron form factors or mass gap of baryon Regge trajectories, other values of $\kappa$ are reported, ranging from $350MeV$ up to $545MeV$ \cite{de2012hadronic,brodsky2008ads}. In \cite{chakrabarti2013generalized} the holographic soft-wall model GPDs, are compared with the phenomenological model of \cite{ahmad2007generalized} and the numerical value of $\kappa$ is obtained by fitting the proton form factors to the experimental data. In this work we use the $\kappa=406MeV$ of Ref. \cite{chakrabarti2013generalized}.

  We continue by introducing the phenomenological model of Ref. \cite{ahmad2007generalized}, where our results will be compared with it. This is a physically motivated parametrization of proton unpolarized GPDs and have the following form:

\begin{equation}\label{phenH}
\begin{split}
   H^q(x,t)&=G_{M_x^q}^{\lambda^q}(x,t) x^{-\alpha^q-\beta_1^q(1-x)p_1t}  \\
   E^q(x,t)&=\kappa_q G_{M_x^q}^{\lambda^q}(x,t) x^{-\alpha^q-\beta_2^q(1-x)p_2t}
    \end{split}
 \end{equation}
For details of functional forms and numerical values of the parameters see \cite{ahmad2007generalized}. The parameters of this model are fixed by fitting the form factors and inclusion of a Regge term for a proper behavior at low $x$.

We finish this section with a discussion of impact parameter dependent GPDs.
It is shown by Burkardt \cite{burkardt2003impact}, that for GPDs at $\xi=0$,  a density interpretation is obtained in the mixed representation of longitudinal momentum and transverse position in the infinite momentum frame \cite{diehl2005generalized}. GPDs in impact parameter space (i.e. Impact parameter dependent PDFs) is obtained by:

 \begin{equation}\label{impact}
q_v(x,\textbf{b})=\int\frac{d^2\Delta}{(2\pi)^2} e^{-i\textbf{b}\Delta}H_v^q(x,t=-\Delta^2)
 \end{equation}
which gives the probability to find a quark with longitudinal momentum $x$ and impact parameter $\textbf{b}$ minus the corresponding probability to find an anti quark \cite{diehl2005generalized}. In Eq.(\ref{impact}), the impact parameter $\textbf{b}$ is the transverse distance between struck parton and the center of momentum (center of momentum is like center of mass but with momentum fractions instead of masses) of the hadron. So Fourier transforming the GPDs (in zero skewness limit) with respect to $t$ gives access to the spatial distributions of partons in the transverse plane. We will show that how the perturbative effects will change the transverse distribution.

\section {Generalized Parton Distributions Evolution}\label{gpdevo}
In this section we explain the factorization of GPDs and its perturbative evolution using the DGLAP equations.
In the DIS formalism imposing the condition that physical quantities must be independent of factorization scale leads to DGLAP equation, which
is the fundamental equation of perturbative QCD. The same factorization procedure occurs for GPDs, and it is the scale which the partons are resolved. We pointed out that in the factorization of exclusive processes to short and long distance , the long distance parts are GPDs and they are evolved also according to the usual DGLAP equations for valence quarks \cite{diehl2005generalized}.
Usually in the literature the scale dependence of the GPDs is implicit in the formalism. The dependence to the factorization scale will be shown explicitly in this section.
In both exclusive and inclusive processes there is a perturbatively calculable part which is scattering of virtual photon from parton, and a nonperturbative uncalculable part which is the rest of the hadron. The scale parameter $\mu$ separates  the perturbative short-distance and nonperturbative long-distance physics. It is a very hard task to determine The scale parameter $\mu$  exactly (like the QCD scale parameter $\Lambda_{QCD}$), but we know that the physical observables must be independent of the choice. The independence from the scale parameter $\mu$, results in the following DGLAP like equation for valence quark GPDs $H_v^q$ (\cite{diehl2005generalized}):

 \begin{eqnarray}\label{dglap}
 \begin{aligned}
\mu^2\frac{d}{d\mu^2}H_v^q(x,t,\mu^2)&=\int_x^1 \frac{dz}{z} [P( \frac{x}{z})]_+H_v^q(z,t,\mu^2)  \\
 &=\frac{\alpha_s(t)}{2\pi}P\otimes H_v^q(t,\mu^2)
 \end{aligned}
 \end{eqnarray}
where $[...]_+$ is the so-called ``+ prescription" for regularization of splitting function and $P(z)=\frac{\alpha_s}{2\pi}C_f \frac{1+z^2}{1-z} $, is the quark splitting function which gives the probability that a quark, having radiated a gluon is left with fraction $z$ of the original momentum. The Evolution equations for $E_v^q(z,t,\mu^2)$, is the same as the Eq.(\ref{dglap}) for $H_v^q(z,t,\mu^2)$.
 At a fixed value of $t$,  Eq.(\ref{dglap})  is a convolution integral that  gives the perturbative changes of GPDs as a function of momentum fraction $x$.
Note that in the usual DGLAP equations there is a term correspond to gluon splitting function, which is absent here because only valence quarks contribute to GPDs sum rules.

 To carry out the calculation of the quark splitting function in Eq.(\ref{dglap}), one needs the coupling constant at different energies, i.e. we need the running of coupling constant in the boundary gauge theory side. From AdS/CFT correspondence we know that the radial coordinate of the bulk theory ($z$) is related to the energy scale of the conformal boundary theory. So a $z$ dependent coupling of the bulk theory is translated to an energy dependent coupling constant of boundary theory. But here we need an exact physical mapping between the gravity and gauge theory sides. Following the method of Ref.\cite{brodsky2010nonperturbative}, light-front holography is used for a precise mapping of the gravity theory coupling in $AdS_5$ to the 4D gauge theory coupling.
To establish the connection, first we should identify the effective coupling of the classical gravity theory. We rewrite the gauge field action of Eq.(\ref{vecaction}) but do not absorb the five dimensional gauge coupling by field strength redefinition:

 \begin{equation}\label{vecaction2}
S_{vec}=\int d^4x dz \sqrt{g}e^{\varphi(z)}\frac{1}{g_5^2}(F^{2})
 \end{equation}
where $\varphi(z)=\kappa^2 z^2$. Now the effective coupling is identified with the prefactor, as:
 \begin{equation}\label{5dcoup}
   g_5^{-2}(z)=e^{\varphi(z)}g_5^{-2}
 \end{equation}
It is known in the context of light-front holography that the $AdS_5$ radial coordinate $z$ is identified with the invariant impact separation variable $\zeta$ of light-front Schrodinger equation \cite{de2009light}. Using the same identification here, we have: $g_5(z) \rightarrow g_{YM}(\zeta) $. Thus :

\begin{equation}\label{alphaads}
 \alpha_s^{AdS}(\zeta)=g_{YM}^2(\zeta)/4\pi \propto e^{-\kappa^2 \zeta^2}
\end{equation}
Now the main step of Ref. \cite{brodsky2010nonperturbative} is to identify the two dimensional Fourier transform of $ \alpha_s^{AdS}(\zeta)$ with respect to $(\zeta,\phi)$, as the physical coupling measured at the scale $Q$, where $\zeta$ and $\phi$ are the two dimensional transverse light-front coordinates. Now  with the ansate that $Q^2$ is the square of the space-like four-momentum transferred to the hadron in the $q^+=0$ light-front frame, Fourier transformation of Eq.(\ref{alphaads}) results in:

\begin{equation}\label{alphaq}
  \alpha_s^{AdS}(Q^2)= \alpha_s^{AdS}(0)e^{-Q^2/4\kappa^2}
\end{equation}
So $\alpha_s^{AdS}(Q^2)$ is the physical nonperturbative running  coupling constant of boundary gauge theory.
The obtained $\alpha_s^{AdS}(Q^2)$ is the same as the QCD coupling constant up to $Q^2\sim 1Gev^2$ but falls-off faster for higher energies. Calculation of quark splitting function in Eq.(\ref{dglap}), using either of $\alpha_s^{AdS}(Q^2)$ and $\alpha_{NLO}^{QCD}(Q^2)$ , gives vanishing contribution above the  $Q^2\sim 1Gev^2$. So their difference is small and we can use either of them to do the calculations.

 Now that we have obtained the coupling constant of the boundary theory, GPDs can be evolve  according to the Eq.(\ref{dglap}). For a fixed value of $t$, it is a convolution integral that gives the evolved GPDs as a function of $x$. Convolution integral can be solved numerically, or by using Mellin transformations.  Mellin transformations convert  the integral equations  into algebraic equations, and after solving the algebraic equations in order to get them back to momentum fraction space $(x)$, an inverse Mellin transform is needed. Both of the mentioned  methods (numerical integration and Mellin transformation) gives  the resulting evolved GPDs as a three dimensional data points (numbers) in $Q$ and $x$ plane. In order to extract the physics , we need the 3D data points to be expressed in terms of functions of $x$ and $Q$. A 3D fitting method is not a good one, because after fitting a function in two variables, there remains some arbitrariness in the functional form. Here we use a lengthy but more accurate method. First at each value of $t$ we fit a proper function of $x$, then doing the same thing in the other way, i.e. for each value of $x$ we fit a proper function of $t$. The calculations are done point by point and the final result is obtained by joining the points.

\begin{figure}
\begin{center}$
\begin{array}{cc}
\includegraphics[width=7cm]{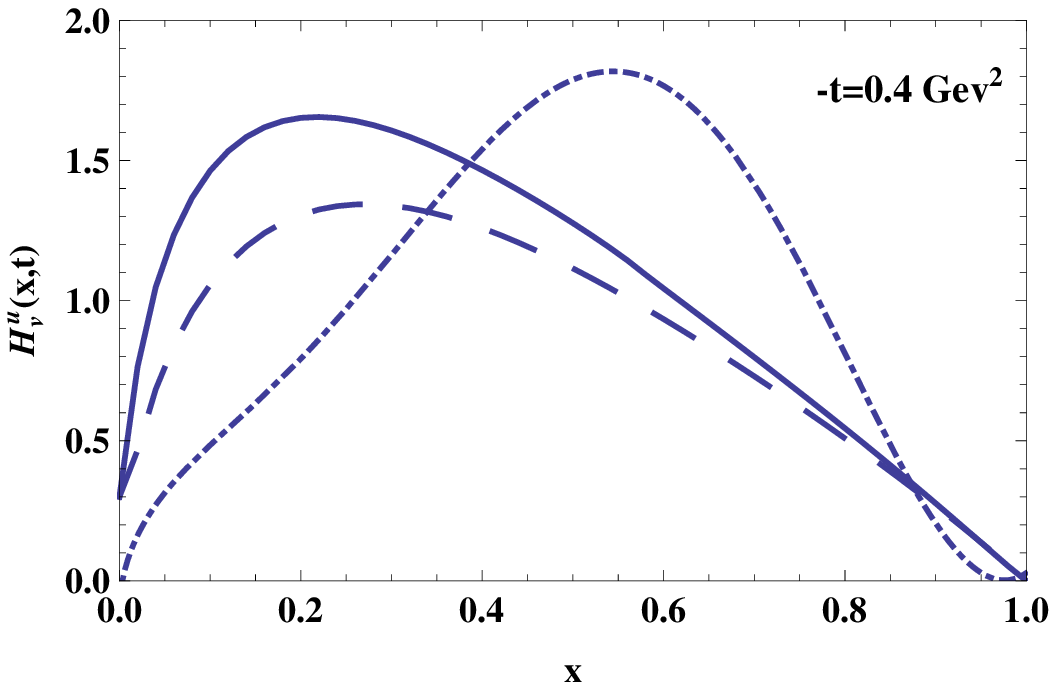} &
\includegraphics[width=7cm]{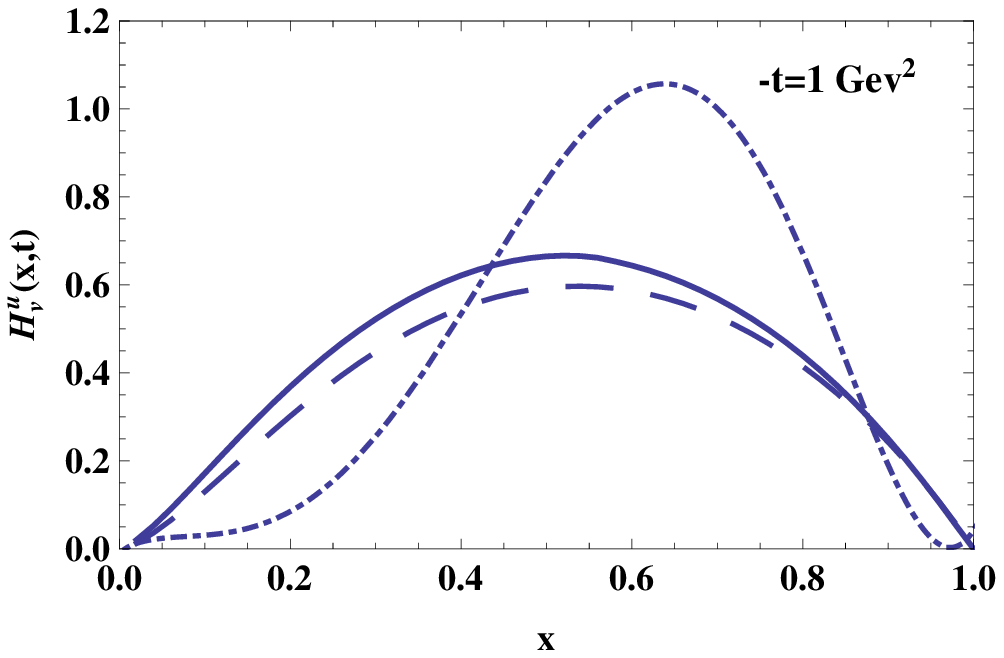} \\
\includegraphics[width=7cm]{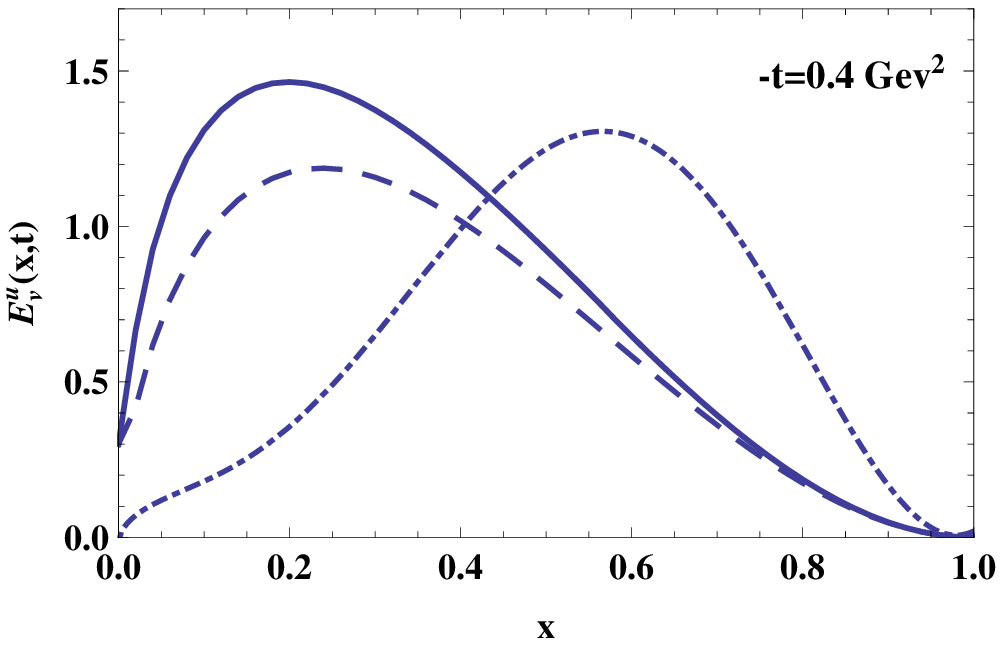} &
\includegraphics[width=7cm]{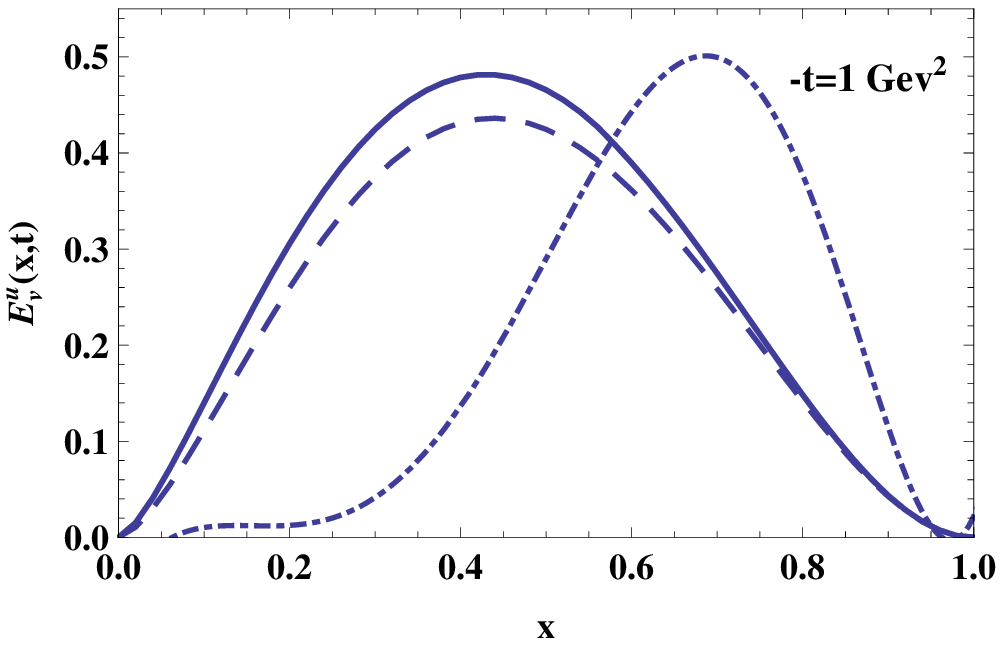}
\end{array}$
\end{center}
\caption{\footnotesize evolved soft wall GPDs (continuous line) compared with original soft-wall GPDs (dashed) and phenomenological model of Ref\cite{ahmad2007generalized}(dashdot) for proton.
The evolved and phenomenological GPDs are of the same order of magnitude at lower values of $-t$. }\label{Huxq}
\end{figure}

First we start to discuss the evolution of GPDs according to Eq.(\ref{dglap}) as a function of $x$ for different fixed values of $t$. In Fig.[\ref{Huxq}] we have shown the original and evolved proton GPDs for two values of $t$ and also compared with the
phenomenological model of Ref. \cite{ahmad2007generalized}, which was introduced in previous section.
 Here some points need to be mention about Fig.[\ref{Huxq}]. Figures show that the evolved GPDs tend to smaller values of $x$, which is the same thing that occurs for PDFs. The reason is that the probability to radiate a gluon is higher for a quark with higher values of momentum fraction, so the distributions incline toward lower $x$ values. The difference of evolved and original soft-wall distributions are bigger for smaller $t$, due to increase in coupling constant and is a little bigger for $E_u(x,t)$ than that of $H_u(x,t)$ at higher energies.

 Note that because of the limitations of perturbative schemes, we can't go arbitrarily down to $Q\rightarrow 0$ region. As discussed in Ref.\cite{gluck1995dynamical}, the perturbative calculations are reliable at most down to $Q^2=0.3GeV^2$.

\begin{figure}
\begin{center}$
\begin{array}{cc}
\includegraphics[width=7cm]{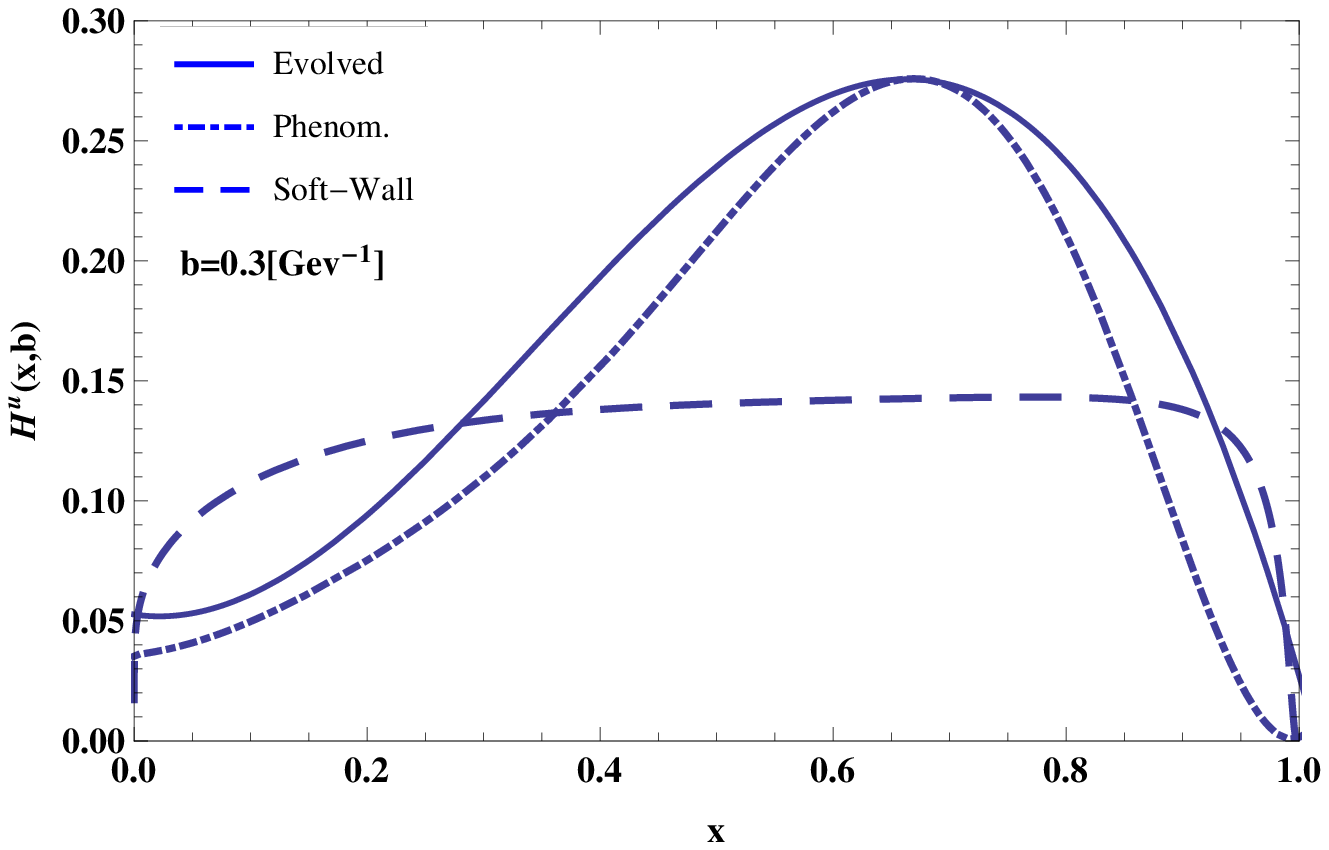} &
\includegraphics[width=7cm]{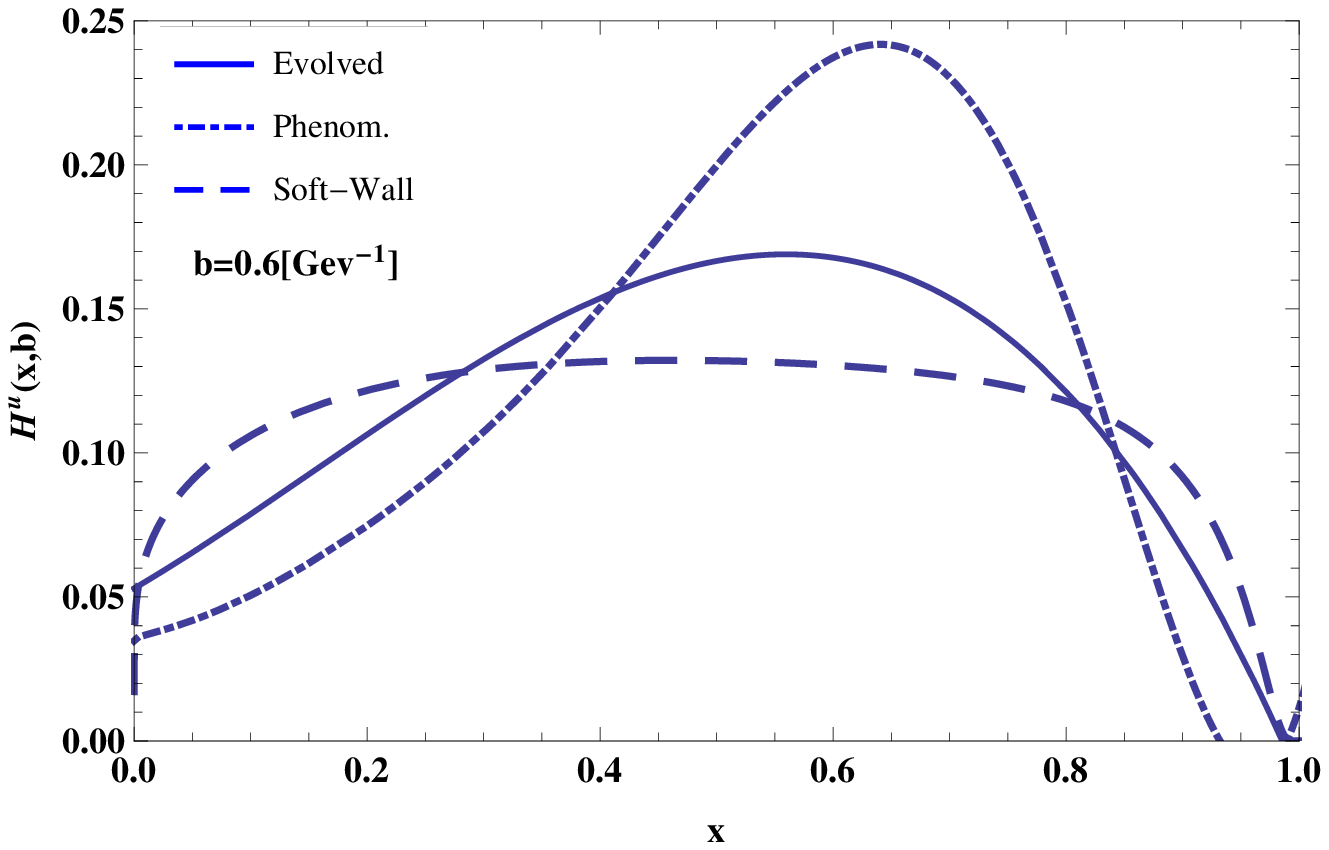} \\
\includegraphics[width=7cm]{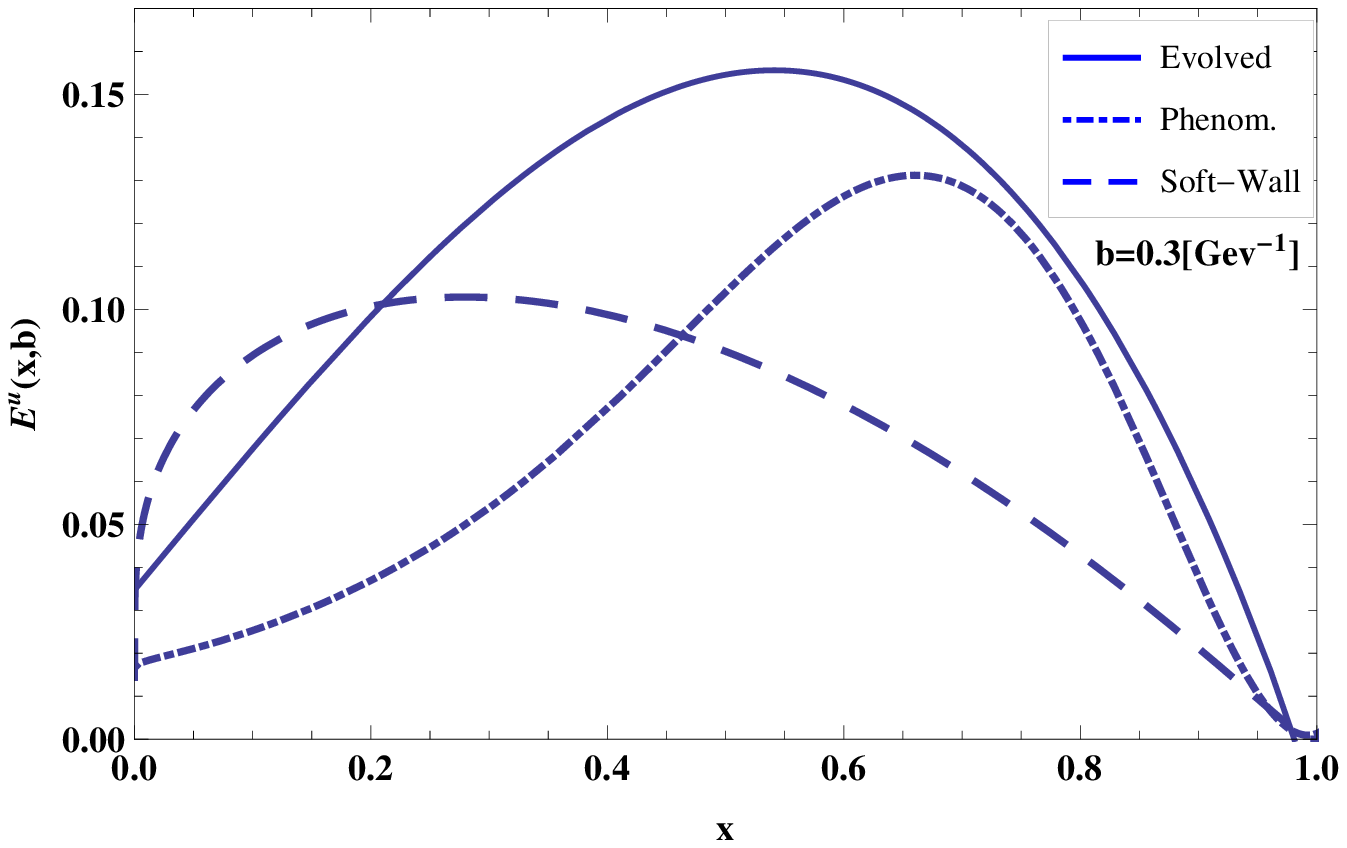} &
\includegraphics[width=7cm]{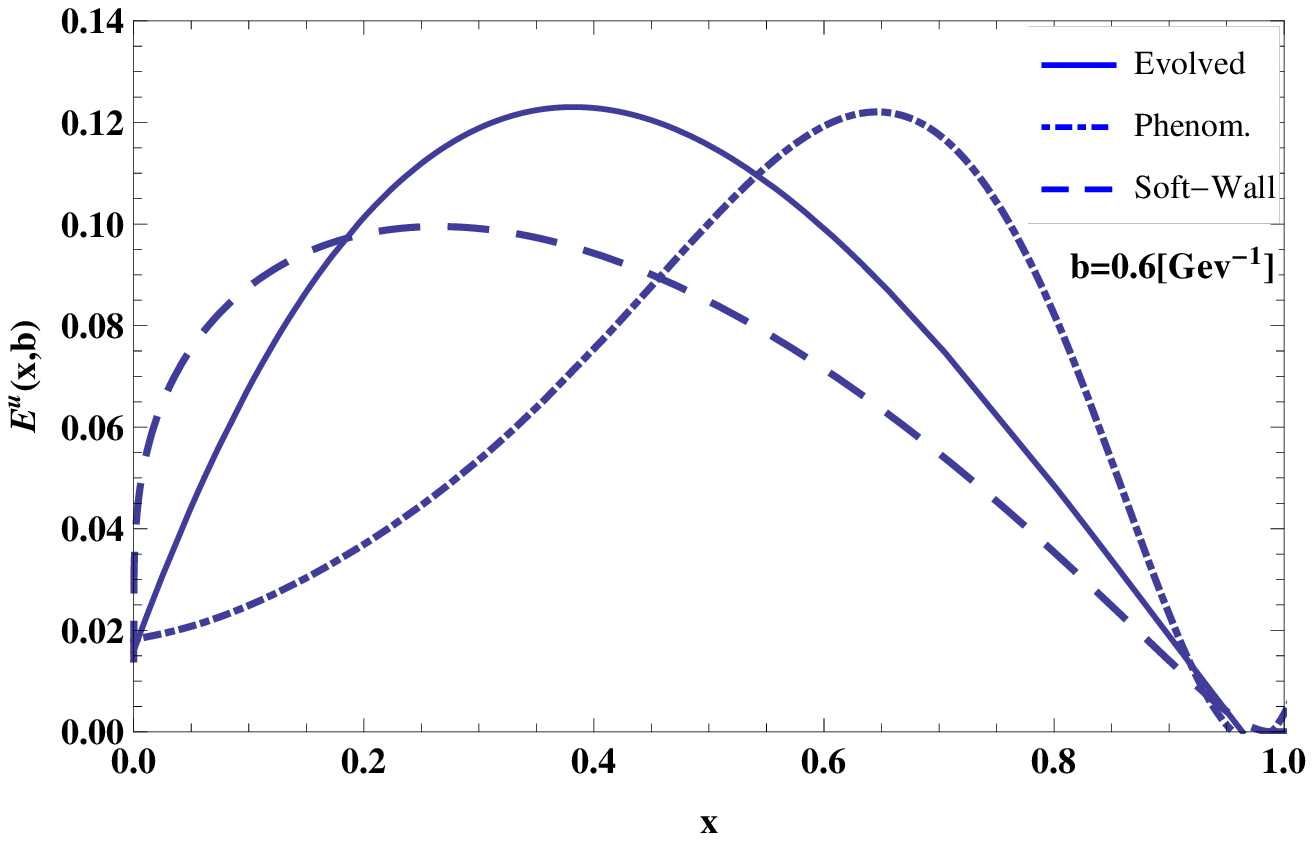}
\end{array}$
\end{center}
\caption{\footnotesize impact parameter dependent proton GPDs in $x$ for two fixed values of $b$. Up: left and right figures are plots for $H^u(x,b)$.
 Down: left and right are plots for $E^u(x,b)$.}\label{2HEbxx}
\end{figure}

Now we investigate the implications of evolution on impact parameter dependent GPDs, which gives much more better results.
 As explained in Sec.\ref{gpdsw}, Fourier transforming the GPDs to impact parameter space, makes it possible to interpret them as distributions. Again we compare the evolved GPDs in impact parameter space with the original and phenomenological ones. Two different cases will be considered for the impact parameter dependent GPDs, first the GPDs will be considered  as  functions of momentum fraction $x$ and fixed value of $b$ as shown in Fig.[\ref{2HEbxx}]. The results are shown for impact parameter dependent proton GPDs $H^u(x,b)$ and $E^u(x,b)$, at two fixed values of $b$.

\begin{figure}
\begin{center}$
\begin{array}{cc}
\includegraphics[width=7cm]{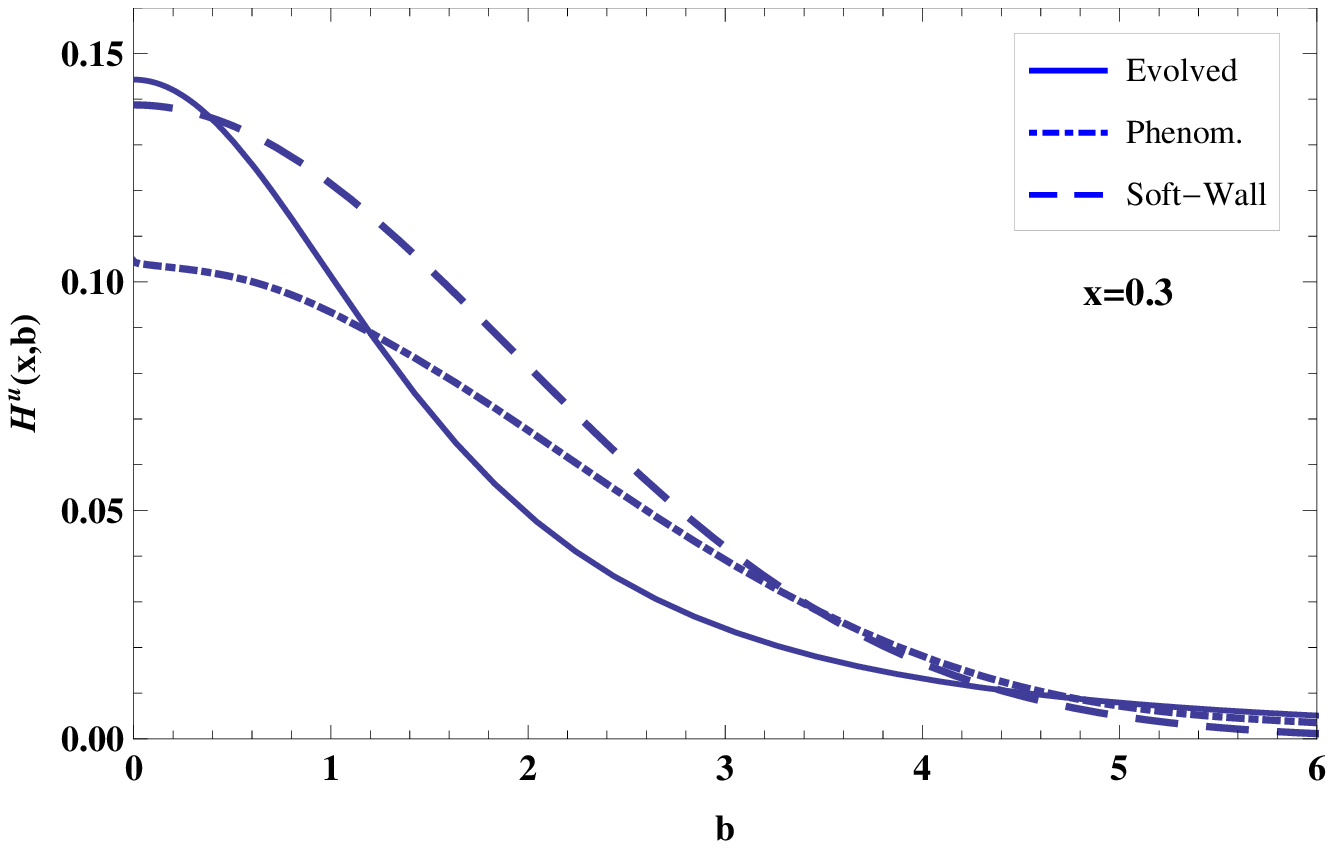} &
\includegraphics[width=7cm]{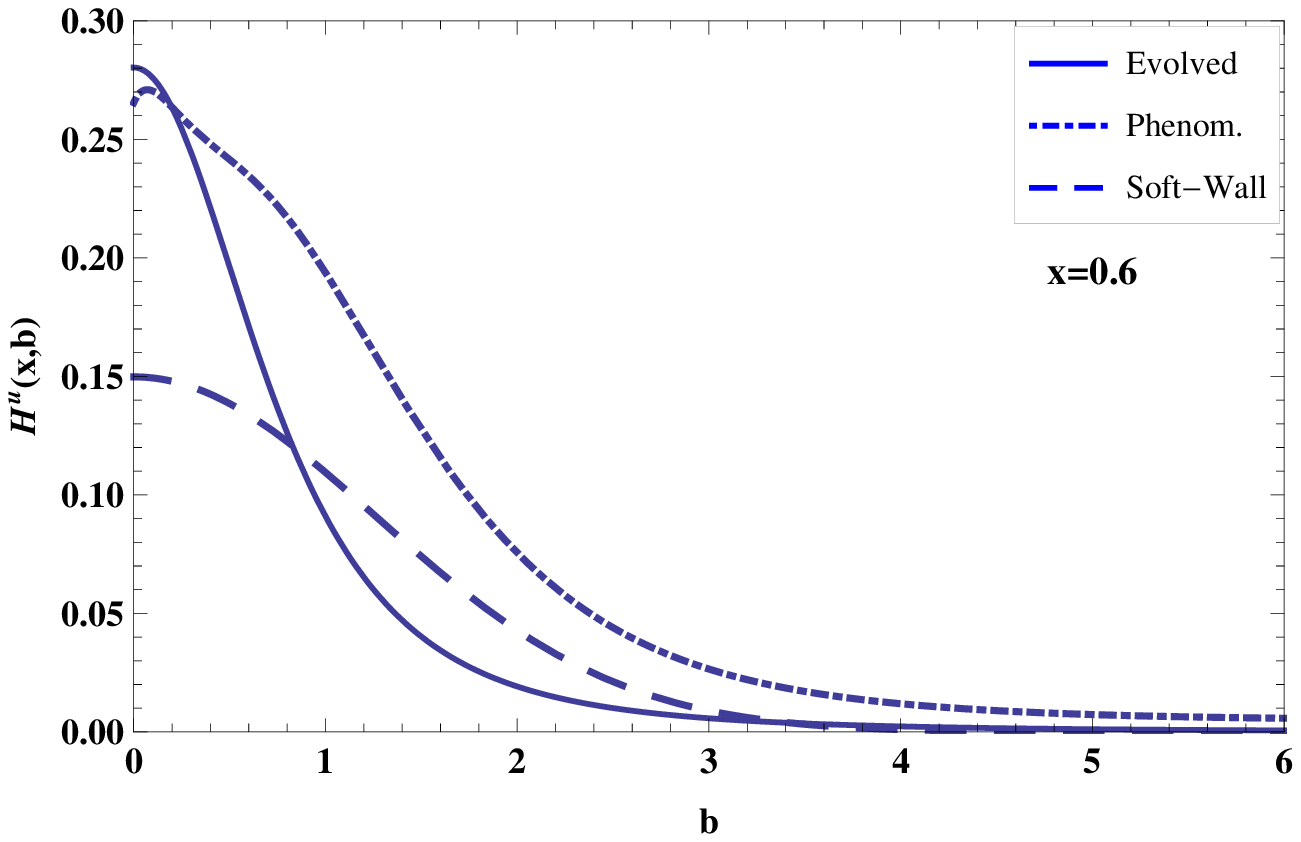} \\
\includegraphics[width=7cm]{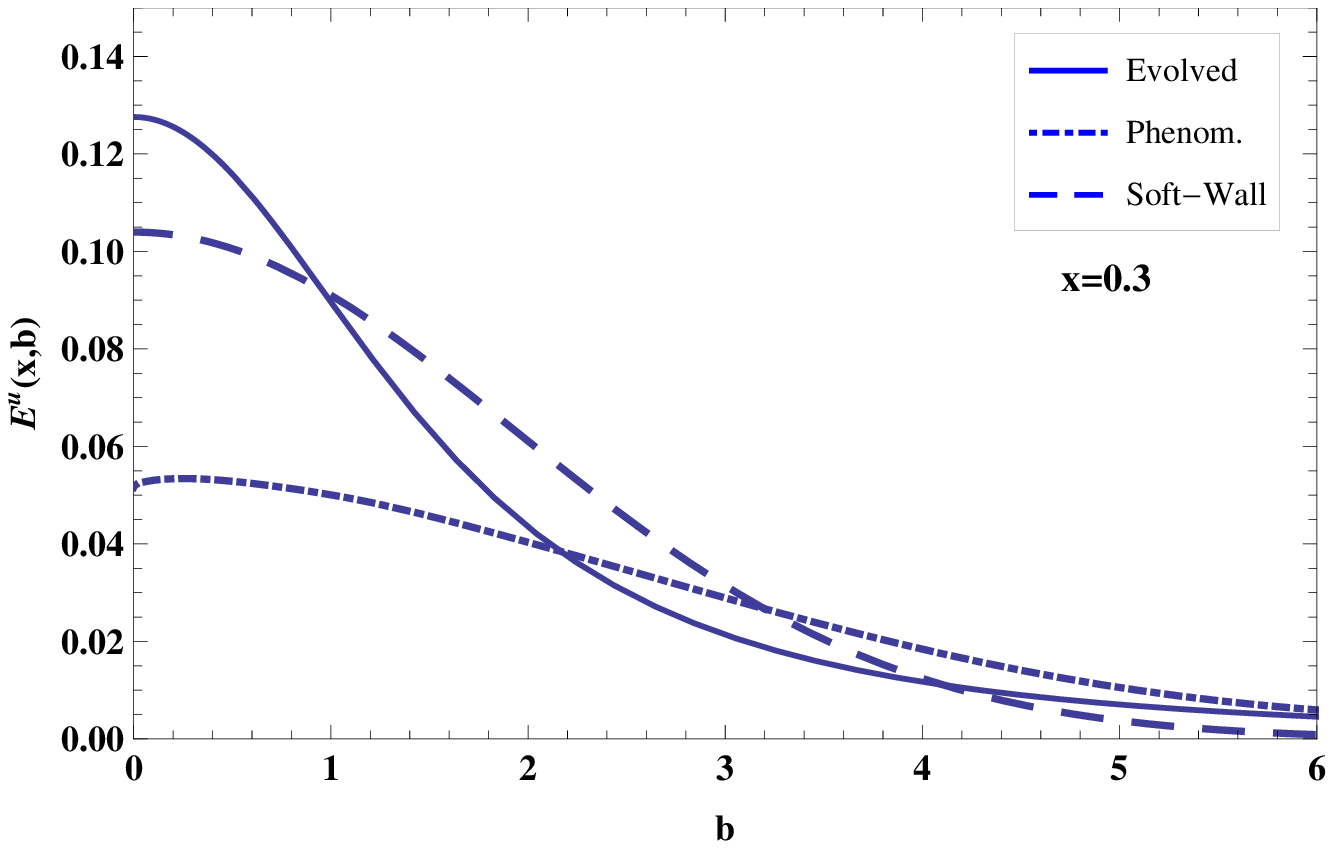} &
\includegraphics[width=7cm]{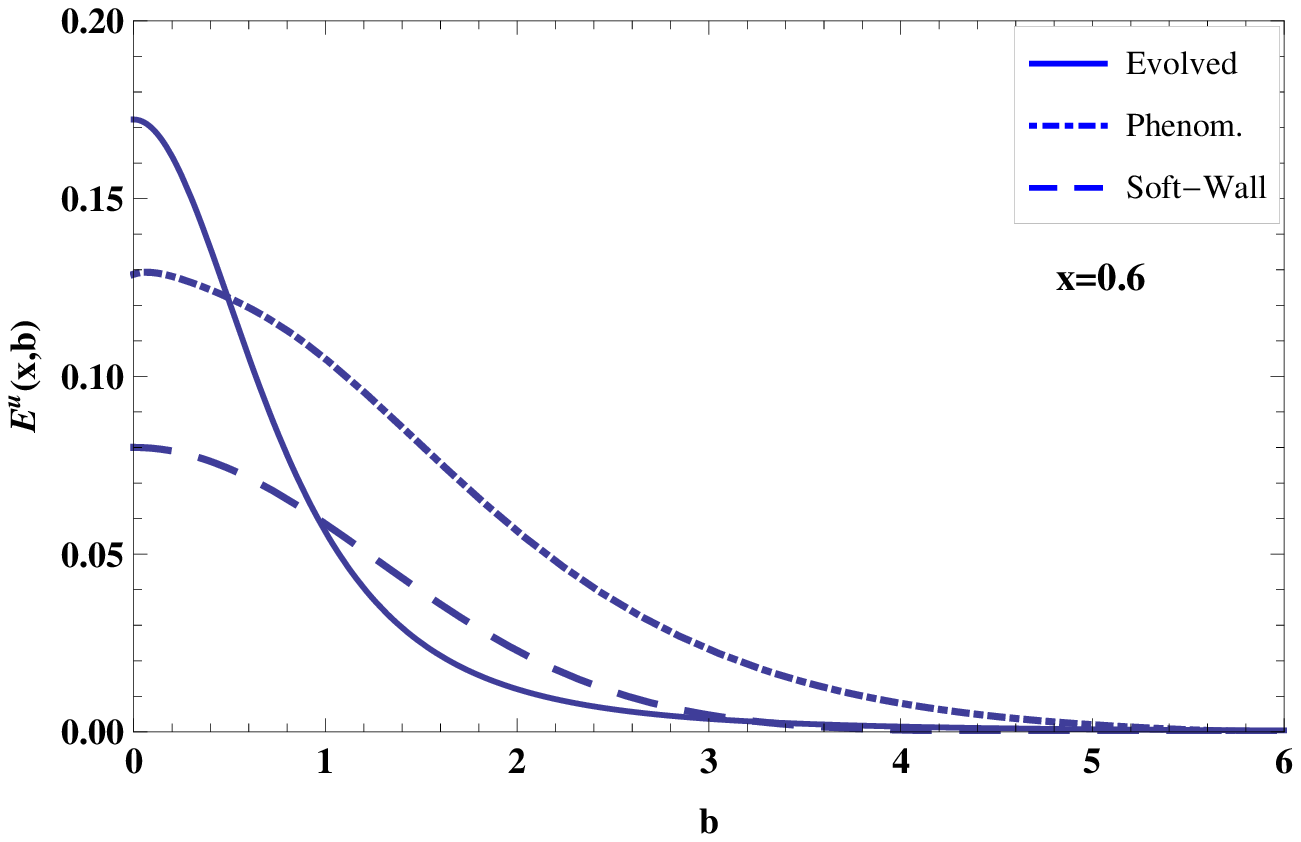}
\end{array}$
\end{center}
\caption{\footnotesize Impact parameter dependent proton GPDs in $b$ for two different fixed values of $x$. Up: left and right are plots for $H^u(x,b)$.
 Down: left and right are plots for $E^u(x,b)$.}\label{3HExbb}
\end{figure}

 In the other case, GPDs are considered as functions of impact parameter $b$ for fixed values of $x$, Fig.[\ref{3HExbb}],and again for the two impact parameter dependent GPDs  $H^u(x,b)$ and $E^u(x,b)$.
 Fig.[\ref{3HExbb}] shows the previously mentioned results in a different way and  that there is something interesting about the evolved distributions. The Inclusion of perturbations make the physics of impact parameter dependent GPDs of holographic soft-wall model compatible with the phenomenology.  It can be seen that the original distributions without any peak or with a peak placed at a different value from the phenomenological model, changes towards the correct physical value.
 Also it can be seen that the mentioned GPDs have different behavior at lower values of impact parameter $b$ and they coincide as $b$ increases.
We can say that for the impact parameter dependent GPDs ,  the evolution causes bigger changes in partonic distributions at smaller values of transverse position parameter $\textbf{b}_\perp$ and except for very small values, it results in a better agreement with phenomenology.
Figures show that at least for limited range of parameters, the physics of proton GPDs is contained in the combined holographic soft-wall model and perturbative effects.

\section {SUMMARY}\label{conclu}
We studied perturbative effects in the context of holographic soft-wall model. We entered the effects through the evolution of GPDs . The coupling constant of QCD like dual theory was obtained using the light-front holography. The GPDs were Fourier transformed to impact parameter space where it is possible to interpret them as distributions. It was shown that how different regions of kinematical variables are affected by perturbations. It was interesting that the combination of holographic soft-wall model and perturbations, results in the correct physics of the GPDs, at least qualitatively.  The method can be used to realize the contribution of evolved GPDs on physical observables, for example through sum rules.
Also it can be applied to the phenomenological model itself for a better fit to experimental data by rearranging the parameters.

\appendix
\numberwithin{equation}{section}

\section {Appendix}\label{appena}
The $C_i(Q^2)$'s in Eq.(\ref{f1f2}), are of the form:
\begin{equation}\label{cci}
\begin{aligned}
    C_1(Q^2)&=\int dze^{-\Phi}\frac{V(Q,z)}{2z^3}(\psi_L^2(z)+\psi_R^2(z))  \\
      C_2(Q^2)&=\int dze^{-\Phi}\frac{\partial_zV(Q,z)}{2z^2}(\psi_L^2(z)-\psi_R^2(z)) \\
       C_3(Q^2)&=\int dze^{-\Phi}\frac{2m_N V(Q,z)}{2z^2}(\psi_L(z) \psi_R(z))   \\
    \end{aligned}
    \end{equation}
where $\psi_L(z)$ and $\psi_R(z)$ are dual to the left-handed and right-handed nucleon fields:
  \begin{equation}\label{psi}
    \psi_L(z)=\kappa^3z^4\;\;,\;\;\psi_R(z)=\sqrt{2}\kappa^2z^3
      \end{equation}
in order to get the GPDs in soft-wall model, one needs the integral representation of bulk-to-boundary propagator \cite{grigoryan2007structure}, given by:

 \begin{equation}\label{v}
    V(Q,z)=\kappa^2z^2\int_0^1\frac{dx}{(1-x)^2}x^{\frac{Q^2}{4\kappa^2}}e^{-\frac{\kappa^2z^2x}{1-x}}
      \end{equation}

\section {Appendix}\label{appenb}

The functional form of $q(x)$ and $\varepsilon^q(x)$ in Eq.(\ref{Hqsw}) are:
\begin{equation}\label{disf}
 q(x)=\alpha^q \gamma_1(x)+\beta^q \gamma_2(x),\;\;\; \varepsilon^q(x)=\beta^q \gamma_3(x)
 \end{equation}
with flavor couplings $\alpha^q$and $\beta^q$ given by:
\begin{equation}\label{11}
\alpha^u=2,\; \alpha^d=1,\; \beta^u=2\eta_p+\eta_n,\; \beta^d=\eta_p+2\eta_n
 \end{equation}
and:

\begin{equation}\label{gamaf}
\begin{aligned}
   \gamma_1(x)&=\frac{1}{2}(5-8x+3x2)   \\
    \gamma_2(x)&=1-10x+21x^2-12x^3      \\
    \gamma_3(x)&=\frac{6\sqrt{2}m_N}{\kappa}(1-x)^2
    \end{aligned}
 \end{equation}
numerical values of parameters are \cite{abidin2009nucleon}:$\kappa=350MeV,\eta_p=0.224$ and $\eta_n=-0.239$.


  \end{document}